\documentclass[11pt,a4paper]{article}
\usepackage{jinstpub}

\usepackage{graphicx}
\usepackage[inline]{enumitem}
\usepackage{siunitx}
\usepackage{textcomp}
\usepackage{hyperref}
\usepackage{tikz}
\usepackage{subcaption}
\usepackage{amsmath}
\usepackage{orcidlink}
\DeclareMathOperator{\median}{median}

\title{A deconvolution based signal reconstruction capable of piled-up
pulse separation}

\author[a, b]{Georgi Georgiev \orcidlink{0000-0001-6884-3942}}
\affiliation[a]{Institute for Nuclear Research and Nuclear Energy, Bulgarian Academy of Sciences, BG}
\affiliation[b]{Physics Faculty, Sofia University ``St. Kliment Ohridski'', BG}
\emailAdd{georgiev.georgi.st@gmail.com}
\arxivnumber{2401.09196}

\abstract{
  This study provides a computationally effective deconvolution algorithm
  capable to reconstruct piled-up events in scintillating detector systems
  with high count rate where fully digitized waveforms are available.  A
  fixed-point iteration algorithm is suggested and used to find properties of
  the signals which are later used during the signal preprocessing stage.  The
  impulse response function is successfully extracted even from heavily
  piled-up event waveforms using an iterative approach.  A methodology for
  pulse time and amplitude reconstruction is based on a deconvolution
  algorithm, which is described in details and some results are presented. The
  presented algorithms are meant to be general and might be successfully
  applied to other fields with minor to no modifications.}

\notoc
\begin{document}

\maketitle
\flushbottom

\section{Introduction}
The present development is inspired by some difficulties in the algorithms for
signal reconstruction of the charged particle vetos system of the PADME
experiment \cite{ref:padme_ven,ref:hindawi_ven}. The veto system is made of
202 plastic scintillating detectors, read out by silicon photo multipliers
(SiPMs) \cite{ref:veto_prototype}. Tailor made controllers operate the SiPM
\cite{ref:sipm_controller} and pass shaped and amplified signals to CAEN V1742
sampling digitizers, responsible for the data acquisition of the experiment
\cite{ref:emanuele_daq}.

Event pile up is not rare in the veto system of the PADME experiment which
makes the reconstruction a non trivial job. An approach based on the exact
impulse response function (IRF) seems inevitable.

Convolution and deconvolution/unfolding algorithms are commonly used in signal
processing and for signals from scintillating detectors in particular. These
algorithms rather appear in the early stages of the processing as a signal
preprocessing or a signal shaping steps \cite{ref:v_yordanov}.

A computationally efficient deconvolution based algorithm is
proposed and studied. The algorithms were initially created with the PADME's
charged particle veto system in mind but they are extendable to other detector
systems where sampled waveforms are available like the small-angle and
electromagnetic calorimeters (SAC and ECal) of the PADME experiment
\cite{ref:sac,ref:ecal}.

Key preprocessing steps like baseline restoration and spike detection are
developed and discussed in the text. The impulse response function is
generally not known, but using the suggested robust iterative procedure it can
be extracted from the experimental data.

The algorithms are tested with some experimental data from the PADME
experiment. Experimental data were also used to obtain the impulse response
function of the detectors which is later used to generate synthetic data for
the survey of the deconvolution based reconstruction.

\section{Signal preprocessing}
The available data from the PADME experiment are sampled at
\SI{2.5}{\giga\hertz} with a \SI{12}{bits} sampling ADC with a dynamic range
of \SI{1}{\volt} which results to a rescaling divisor \SI{4.096}{ADU \per
\milli\volt}.

A not fully explained phenomenon leads to a jump in amplitude in the last few
samples of the waveforms. In order to eliminate possible complications with
the signal reconstruction processing the last 24 samples are substituted with
the mean value of the preceding 10 values.

Amplitudes are further calibrated based on the capacitor performances using
the algorithm provided in \cite{ref:drs4_calibration}.

Additionally custom spike detection and baseline drift calibration algorithms
are applied.

\subsection{Spike detection}
In physics experiments spikes not originating from the studied phenomena are
common. These effects for example could be due to a bit flip or a charge leakage
caused by ionizing particles interacting with the electronic devices of the
data acquisition system. In the experimental data from the PADME experiment
single- and double-bin spikes are present.

The following metrics are used to locate single-bin spikes \(M'\) and for
double-bin spikes \(M''\):

\begin{equation}
M'_i = \left|s_{i+1} - 2\,s_i + s_{i-1} \right| - \left|s_{i+1} - s_{i-1} \right|
\end{equation}

\begin{equation}
M''_i = \left | s_{i+2}- s_{i+1} - s_i + s_{i-1}\right| - \left | s_{i+2}-
  s_{i+1} + s_i - s_{i-1}\right|,
\end{equation}
where \(s_i\) is the amplitude of the \(i\)-th sample. The criterion for a
spike at \(i\)-th sample is:

\begin{equation}
M_i > 12\,\textrm{MAD}(M),
\end{equation}
where \(MAD(M)=\median\limits_{k} \left(\left|M_k -
\median\limits_{l}(M_l)\right|\right) \) is the median absolute
deviation of the metric and the threshold 12 is selected experimentally.

The sensitivity of these simplistic metrics can be affected if the spikes are
located in a sloped section of the waveform but the combination of the two
metrics \(M'\) and \(M''\) provide satisfactory criteria for the purpose of
this analysis.

A deconvolution based reconstruction will be relatively insensitive to
sporadic spikes as they differ significantly from the impulse response
function. On the other hand spikes will be particularly harmful during the
process of IRF estimation. For this reason during the IRF estimation events
with spikes have been discarded. During the signal reconstruction it is
recommended to substitute the corrupted values by linear interpolation of
valid neighbouring samples.

\subsection{Baseline drift}
\begin{figure}[b]
  \begin{center}
    \includegraphics[width=.75\textwidth]{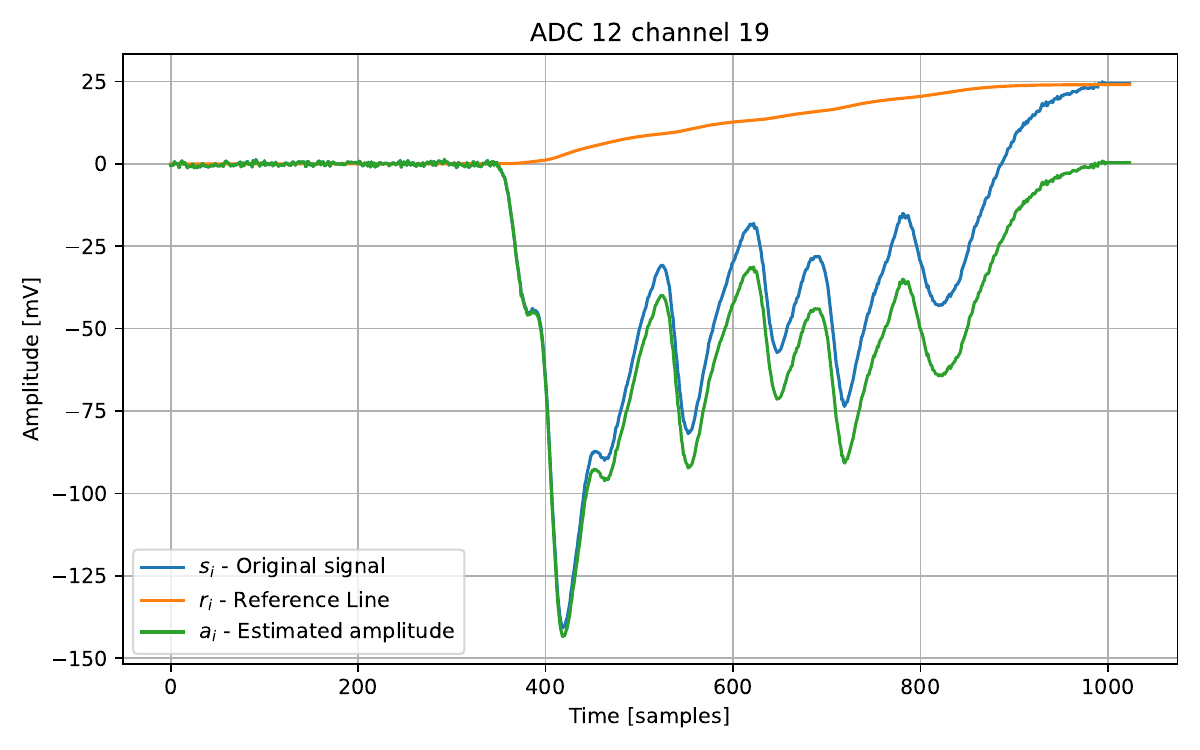}
  \end{center}
  \caption{\label{fig:pedestal_drift} Illustration of baseline drift
  calibration.}
\end{figure}

The experimental data show an effect caused by the SiPM controller:
waveforms with relatively big number of pulses present noticeable drift of the
baseline which seems proportional to the cumulative amplitudes (charge) of the
waveform. Figure \ref{fig:pedestal_drift} illustrates the baseline
drifting effect.
This effect is a problem for both IRF estimation and the deconvolution as they
assume only sum of scaled and time-shifted IRFs and infinite signals so the
first and last value should be equal.

The adopted model for this effect is that the change in the reference \(r\)
depends on the actual real amplitude \(a\) multiplied with a constant
coefficient \(c\), formally defined using the following difference equation:
\begin{equation}
  r_{i} - r_{i-1} = c\,\frac{\overbrace{s_i-r_i}^{a_i}
  +\overbrace{s_{i-1}-r_{i-1}}^{a_{i-1}}}{2}
\end{equation}
The only known value is \(s\), i.e. the signal measured by the ADC. The constant
\(c\) is assumed to be equal for all the channels as the front-end
electronics responsible for this effect is exactly the same for all veto
channels. It is further assumed that:
\begin{equation}
  \label{eqn:ped_drift_assumption1}
r_0 = s_0 = 0
\end{equation}
\begin{equation}
  \label{eqn:ped_drift_assumption2}
  R:=r_{1023} = s_{1023}
\end{equation}
The assumption (\ref{eqn:ped_drift_assumption1}) is valid as the discharge time
constant of the front-end electronics is smaller than event time period.  The
assumption (\ref{eqn:ped_drift_assumption2})
translates to corrected amplitude being zero at the end of the digitization
window. \(R\) is used as shorthand for the last value for the baseline.  This
assumption, however, is not always satisfied because the relatively long
signals from the veto system can go out of the digitization window. The
requirement:
\begin{equation}
  \label{eqn:ped_drift_requirement}
\min\limits_{i\in[800, 1023]} a_i<-5mV
\end{equation}
is believed to filter-out all such events.

The constant \(c\) is estimated using the following recurrent algorithm based on
fixed-point iteration method. Every new approximation \(c_{new}\), based on the
previous one \(c_{prev}\), is defined using the recurrent relation:

\begin{equation}
c_{new} = \frac R
  {\sum\limits_{i=0}^{1023} s_i - \sum\limits_{i=1}^{1023} \frac{\frac
  {c_{prev}} 2 s_i +\frac {c_{prev}} 2 s_{i-1} + (1 - \frac {c_{prev}} 2)
  r_{i-1}} {1+ \frac {c_{prev}} 2}}
\end{equation}

The provided relation tends to overshoot the correct value so the convergence
has been accelerated using:

\begin{equation}
c_{new;\,acc} = \frac 3 4 c_{new} + \frac 1 4 c_{prev}
\end{equation}

Multiple events have been analysed. For each event the constant \(c\) was
calculated using the fixed-point iteration. The global value of the constant
\(c\) across different channels has been found to be:
\[c_{best} = -0.0007134(97).\]

\section{Signal reconstruction}
A computationally cheap algorithm that is able to locate in time and to find
the amplitudes of pulses in digitized waveforms is needed.  Even though Wiener
deconvolution with some kind of regularization like the one provided in
\cite{ref:sparce_deconvolution} is possible, it was considered not applicable
due to the big amount of data that needs to be processed and the computation
cost of the algorithms.  An heuristic method for the deconvolution is
suggested and tested.

\subsection{Impulse response function} \label{sec:impulse_response_function}
The only available information is the digitized waveforms. No theoretical or
other a priori estimates for the impulse response functions or the positions
and amplitudes of the pulses are available. The impulse response function
needs to be extracted from the multiple events with overlapping pulses using
iterative approach.

\subsubsection{Initial guess}
The iterations start from an initial guess defined as:
\begin{equation}
h_0(t) = \textrm{median}\frac{A_i(t-\arg \min\limits_\tau A_i(\tau))}{\min\limits_\tau A_i(\tau)},
\end{equation}
where $A_i(t)$ is the $t$-th sample of the $i$-th signal.  This can be
visualized as shifting the waveforms so that their absolute minima are located
at zero time position, then calculate median values for each time position
across all waveforms.  The motivation is that only the signals from pulses with
the biggest amplitude will correlate as they are forced to be at same time
position.  All other signal from pulses with smaller amplitudes will appear at
random positions. By taking median average the initial approximation for the
impulse response function will appear at zero time position and the
contribution of the random pulses will diminish.

\subsubsection{Iterations} Assume
\begin{equation}
s = d \ast h + \varepsilon,
\end{equation}
where \(s\) is the digitized waveform, \(d\) is a sparse vector having the
pulse amplitudes the corresponding time positions, \(\ast\) denotes
convolution, \(h\) is the impulse response function and \(\varepsilon\) is
additive noise.

The estimated pulses with the current approximation of the impulse response
function will be:
\begin{equation}
  d_{i, new} =deconvolve(s_i, h_{prev}),
\end{equation}
where \(i\) denotes the sequential number of the waveform and \(deconvolve\)
is the deconvolution method described in section
\ref{sec:deconvolution}.  Then the new approximation for the impulse response
function derived from the \(i\) waveforms will be:
\begin{equation}
  h_{i, new} = s_i \ast d^{-1}_{i, new},
\end{equation}
where \(d^{-1}\) is the inverse of \(d\) in sense \(d^{-1}=\mathcal{F}^{-1}\left({1}/{\mathcal{F}(d)}\right)\).

The differences in pulse amplitudes require a maxima based scaling:
\begin{equation}
  \bar h_{i, t} =\frac{ h_{i,t}}{  \max\limits_t h_{i, t}}
\end{equation}

At the end of each iteration an improved version of the impulse response
function is calculated as:
\begin{equation}
  \hat h_{new} = \mathrm{median}(\bar h_{i, new}).
\end{equation}

\begin{figure}[h]
  \begin{subfigure}[t]{.48\textwidth}
    \centering
    \includegraphics[width=\textwidth]{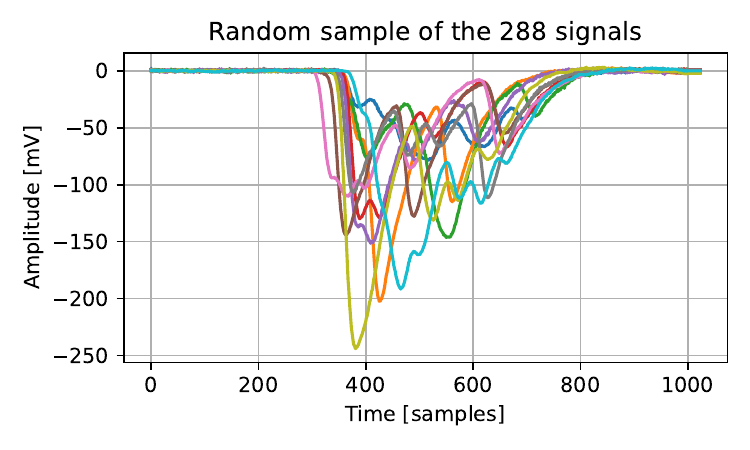}
    \caption{\label{fig:irf_iter_signals} Waveforms used to test the
    robustness of the algorithm. The pulses in all waveforms overlap
    making it difficult to extract the impulse response function.}
  \end{subfigure}
  \hspace*{\fill}
  \begin{subfigure}[t]{.48\textwidth}
    \centering
    \includegraphics[width=\textwidth]{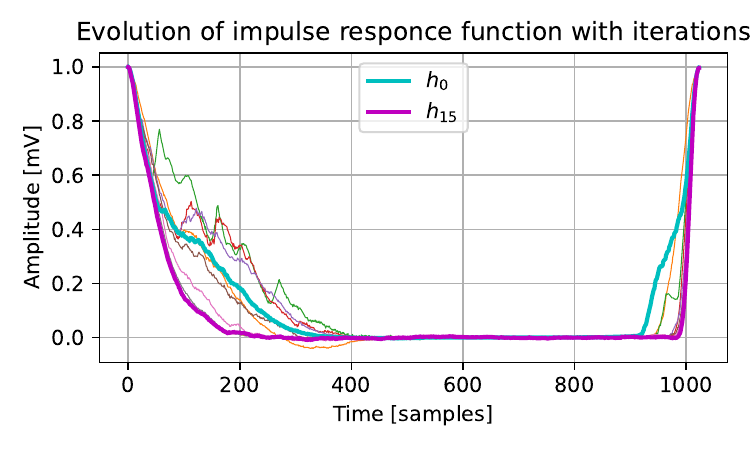}
    \caption{\label{fig:irf_iter_evolution} The evolution of the estimates for
    the impulse response function across the iterations is visualized. The
    initial guess and the final result are shown with ticker lines.}
  \end{subfigure}

  \caption{\label{fig:irf_iter} Illustration of the robustness of the
  iterative procedure using 288 piled-up events.}
\end{figure}

The robustness of the iterative procedure is demonstrated using only a limited
number of piled-up events sample of which is shown on figure
\ref{fig:irf_iter_signals}. Even with the absence of a clear picture of the impulse
response function within 15 iterations a good estimate was found. The
evolution of the IRF estimate is shown on \ref{fig:irf_iter_evolution}.

\subsection{Deconvolution} \label{sec:deconvolution}

\subsubsection{IRF inversion}

\begin{figure}[t]
  \begin{center}
    \includegraphics[width=.75\textwidth]{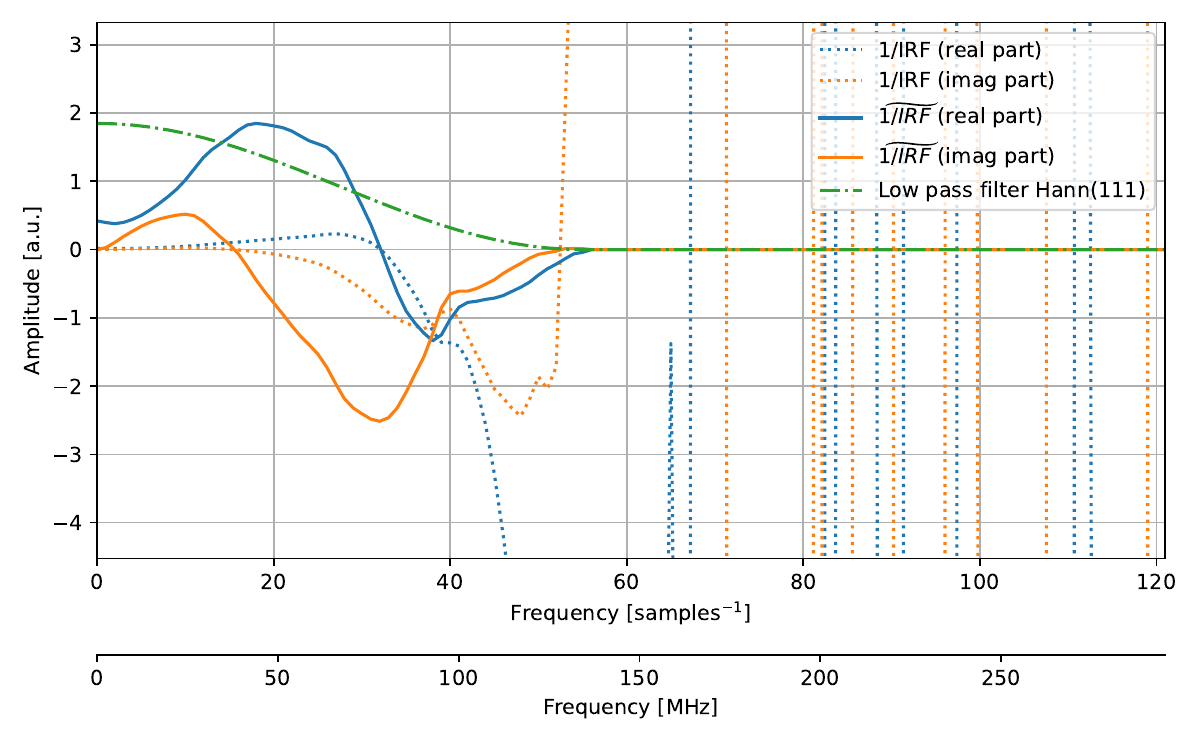}
  \end{center}
  \caption{\label{fig:irf_lowpass} The real and imaginary part of the inverse
  of the impulse response function \(h^{-1}\) are shown with dotted line.
  The dash-dotted line shows the spectrum of the low-pass filter. The solid
  lines represent the inverse of the impulse response function with the
  low-pass filter applied \(\widetilde{h^{-1}}\).}
\end{figure}

The impulse response function \(h\) is a smooth and noiseless function. Taking
its inverse \(h^{-1}\) results in huge amplification in the high frequencies,
which is shown in frequency domain on figure \ref{fig:irf_lowpass} with dotted
lines.  Applying a low-pass filter will remove the high frequency noise at the
price of broadening the result of the deconvolution. So instead of sharp lines
at the pulse position, smooth symmetrical shapes are expected. This will require
further processing. The result with applied low-pass filter
\(\widetilde{h^{-1}}\) is shown on figure \ref{fig:irf_lowpass} with solid
lines.

The spectrum of the estimated impulse response function was studied and it was
found that the noise becomes dominant after the 56-th frequency bin.

Low-pass filters based on Blackman-Harris and Hamming windows have been tested
with the idea to achieve minimal leakage and narrower main lobe
correspondingly. A better result, however, was obtained by applying Hann
window corresponding to cut-off frequency \SI{50}{\mega\hertz} at
\SI{3}{\deci\bel} illustrated with the dash-dotted line on figure
\ref{fig:irf_lowpass}. The estimated cut-off frequency is in agreement with
the \SI{70}{\mega\hertz} bandwidth of the amplifier in the SiPM controllers
\cite{ref:sipm_controller}.
It is practical to store the inverted impulse response
function multiplied with the low-pass filter \(\widetilde{h^{-1}}\) as it has
only 56 non-zero complex numbers and can be directly used in the processing of
new events by complex multiplication in the frequency domain.

To facilitate the comparison with other existing algorithms which are using
maximal value of the pulse as the pulse amplitude (as opposed to pulse shape
integral), \(\widetilde{h^{-1}}\) is normalized so that
\begin{equation}
  \max\limits_t \left(h \ast \widetilde{h^{-1}}\right)_t = 1.
\end{equation}
As the deconvolved signal
\begin{equation}
  \widetilde d = s \ast \widetilde{h^{-1}}
\end{equation}
is a result of a shallow deconvolution (a low-pass filter is applied to obtain
$\widetilde{h^{-1}}$), the information about position in time and amplitude
will be encoded in rather broader symmetric shapes.

\subsubsection{Pulse estimate}
\begin{figure}[h]
  \begin{center}
    \includegraphics[width=.75\textwidth]{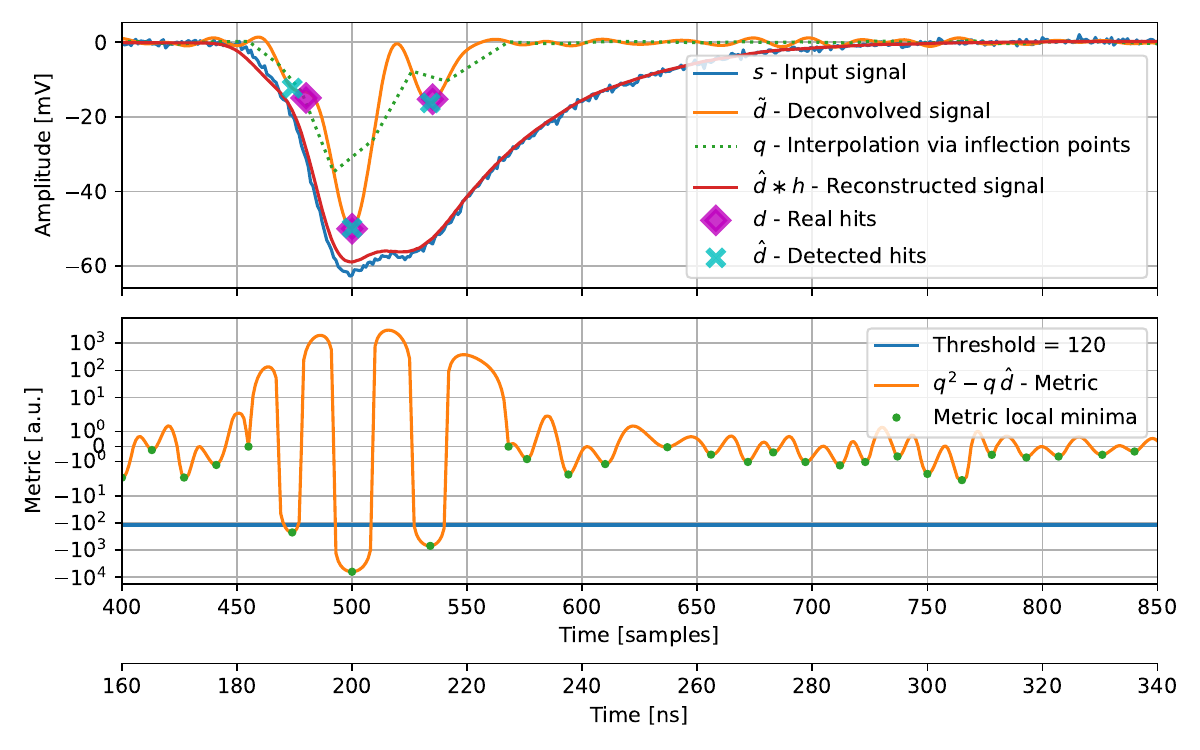}
  \end{center}
  \caption{\label{fig:deconvolution_example} Deconvolution example using
  artificially generated signal. True and reconstructed signals are plotted
  together with other functions involved in the deconvolution process.}
\end{figure}

The Hann window leakage leads to oscillations in the deconvolved signal
\(\widetilde d\),
which are visible on figure  \ref{fig:deconvolution_example}. These fluctuations
are around the true value, i.e. around zero in absence of pulses and non-zero in
presence of pulses.  A linear interpolation via the inflection points of the
deconvolved signal can be used as a weighting factor. The relative weight of
the regions with pulses will be increased while the oscillations in ranges without
pulses get suppressed.  This heuristic metric is defined as
\begin{equation}
  M = q^2 - q\, \widetilde{d},
\end{equation}
where \(q\) is the linear interpolation between the inflection points of the
deconvolved signal \(\widetilde{d}\). The inflection points are found as the
positions of all local extrema of the successive differences
of \(\widetilde{d}\).

The positions of the minima of the metric \(M\) are candidates for pulse
positions. Resulting metric values for all pulse candidates are normalized
using median absolute deviation. This helps distinguish real events from the
noise. A median value, which is computationally cheaper, can also be used for
normalization with similar outcome.

At the moments where the minima of the metric are below the
threshold of \(-120\) (which was found to produce good results with the
given set of experimental data) pulses are declared.
The final result \(\widehat d\) (the estimate for \(d\)) is a sparse vector
with values equal to the amplitudes of the deconvolved signal
\(\widetilde{d}\) at the corresponding time positions.

\section{Performance and discussion}

\begin{figure}[!h]
  \centering
  \begin{subfigure}[t]{.75\textwidth}
    \centering
    \includegraphics[width=\textwidth]{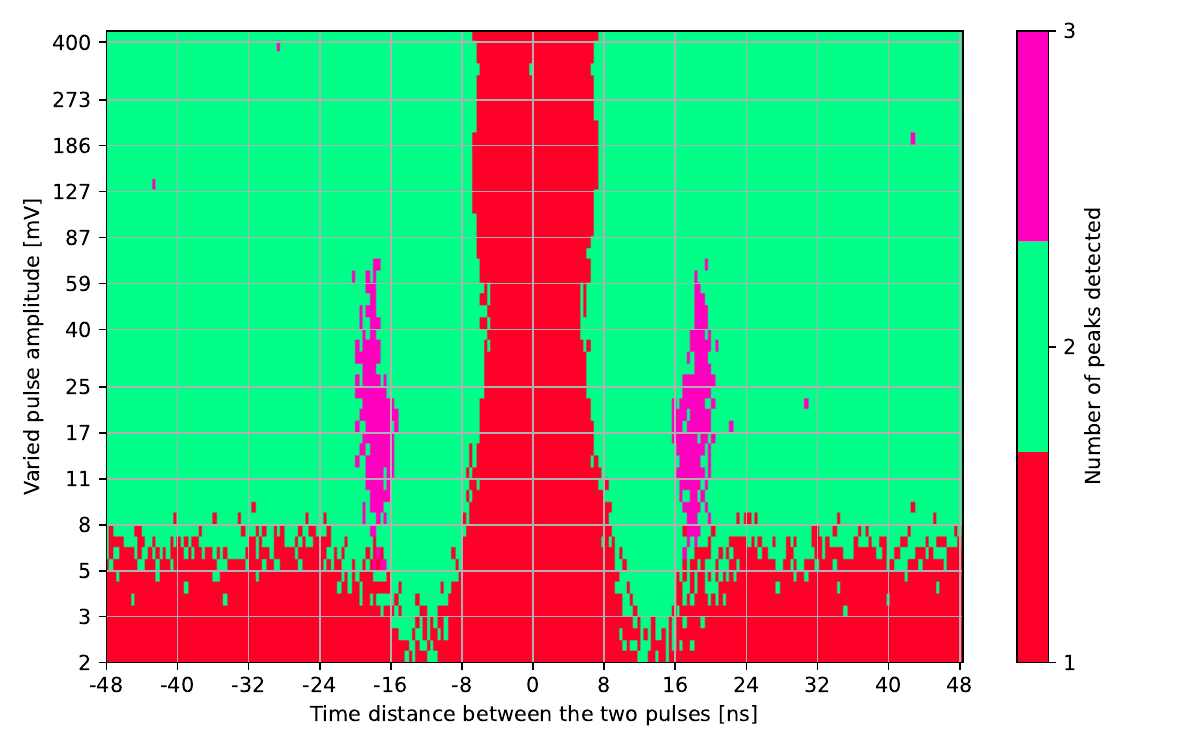}
    \caption{\label{fig:scan_peak_counts} The map shows the detected number
    of pulses for each pair amplitude / time offset.}
  \end{subfigure}
  \begin{subfigure}[t]{.48\textwidth}
    \centering
    \includegraphics[width=\textwidth]{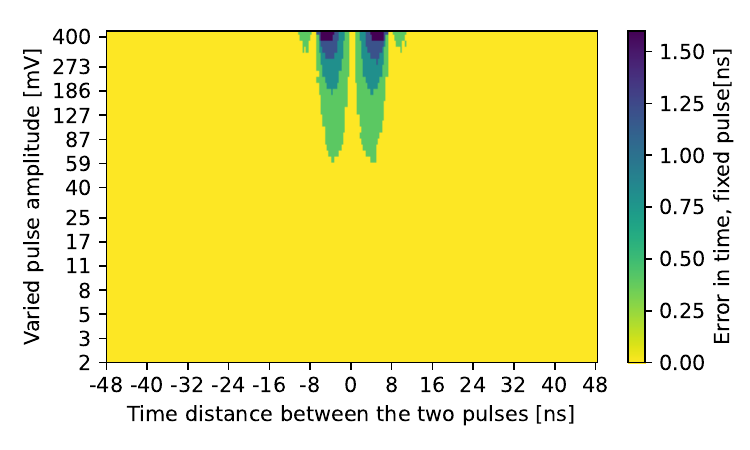}
    \caption{\label{fig:scan_err_time1} Errors in the reconstructed time of
    the fixed pulse.}
  \end{subfigure}
  \hspace*{\fill}
  \begin{subfigure}[t]{.48\textwidth}
    \centering
    \includegraphics[width=\textwidth]{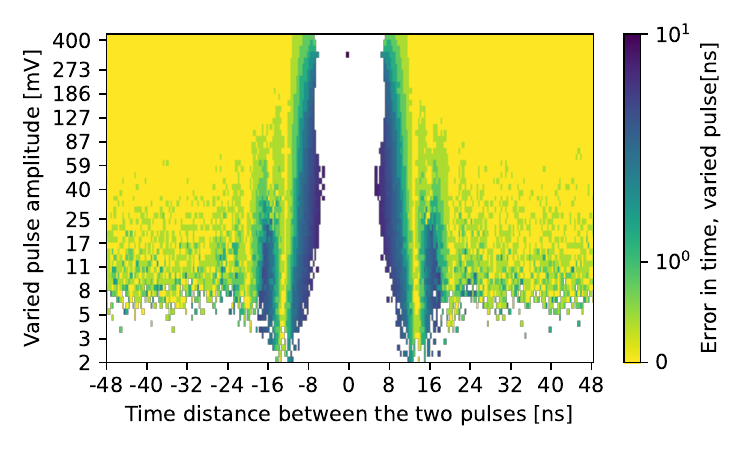}
    \caption{\label{fig:scan_err_time2} Errors in the reconstructed time of
    the varied pulse.}
  \end{subfigure}

  \begin{subfigure}[t]{.48\textwidth}
    \centering
    \includegraphics[width=\textwidth]{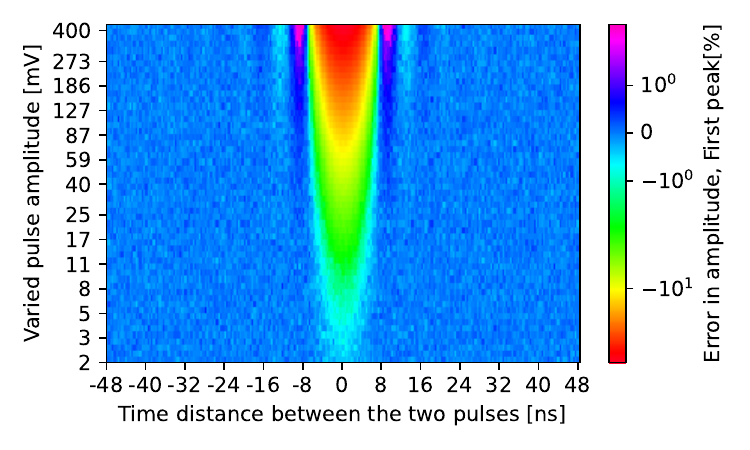}
    \caption{\label{fig:scan_err_ampl1} Errors in the reconstructed amplitude
    of the fixed pulse.}
  \end{subfigure}
  \hspace*{\fill}
  \begin{subfigure}[t]{.48\textwidth}
    \centering
    \includegraphics[width=\textwidth]{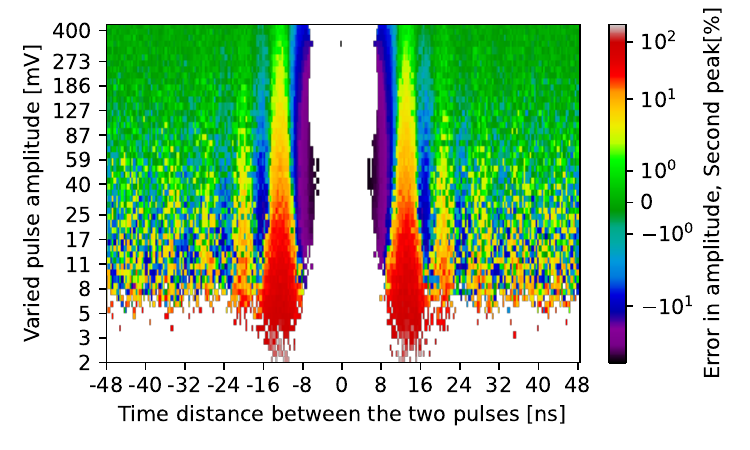}
    \caption{\label{fig:scan_err_ampl2} Errors in the reconstructed amplitude
    of the varied pulse.}
  \end{subfigure}

  \caption{\label{fig:scan} Performance of the proposed reconstruction
  algorithm using artificially generated events. The generated waveforms have
  exactly two signals: one fixed at position and amplitude
 and a second with variable position and amplitude}
\end{figure}

For better understanding of the behaviour of the reconstruction algorithm, the
phase space was sampled using a signal generator. The generated waveforms
always have two pulses with the calculated IRF using the iterative algorithm
described in \ref{sec:impulse_response_function}. A fixed pulse is set in
position 500 and amplitude \SI{550}{\milli\volt} and a variable pulse is set
with amplitude in the range $A\in[2, 400]\,\si{\milli\volt}$ and time in the
range \(t_{offset}\in[380, 620]\), (which translates to
\SI{\pm48}{\nano\second} time difference with respect to the fixed pulse).
Gaussian noise with \(\sigma = \SI{0.55}{\milli\volt}\) is added to the
signal. The \(\sigma\) value is obtained from experimental data.  The
synthetic test signal with two pulses, resembling the experimental waveforms
from the PADME veto system, is defined as:
\begin{equation}
s[t] = 550\,h[t-500]+A\,h[t-t_{offset}] + \mathcal{N}(0, 0.55).
\end{equation}

The test signals are processed with the presented algorithm to study its
performance. Figure \ref{fig:scan_peak_counts} shows the number of
reconstructed pulses in the generated synthetic waveforms. As the waveforms
have exactly two pulses, a single reconstructed pulse corresponds to a false
negative, while the three-pulse case corresponds to a false positive.  An
explanation for this behaviour can be found in figure
\ref{fig:deconvolution_example} bottom, where the metric \(M\) is shown.
Every peak corresponding to a pulse is surrounded by two deep valleys which
originate from the side lobes caused by the low-pass filter.  Thus if a varied
pulse occurs within range \SI{\pm8}{\nano\second} around the fixed pulse, it
remains undetected, as it will not be able to pass the threshold in \(M\).
Pulses also cannot be detected reliably when their amplitude is smaller than
\SI{8}{\milli\volt}. A pulse located just next to a valley in \(M\) caused by
a bigger pulse, tends to be detected as two pulses.  The algorithm show a
dead-time window which is \SI{\pm8}{\nano\second} wide.  The effect of
detecting a single pulse as two would not lead to significant inefficiency of
the PADME veto system as the detected pulses are usually about
\SI{1}{\nano\second} apart.

Figures \ref{fig:scan_err_time1} and \ref{fig:scan_err_time2} show the
absolute errors of the reconstructed time for the fixed and for the varied
pulses. It is visible that the variable pulse can causes error in the
reconstructed time of the fixed pulse. The deviation from the true value is
within \SI{\pm1.5}{\nano\second} and appears only when pulses with similar
amplitudes are less than \SI{8}{\nano\second} apart. The reconstructed time
for the varied pulse shows errors of about few \si{\nano\second} when close to
the fixed one. This behaviour again is explained with valleys that appear in
the metric \(M\).

The errors in amplitudes are shown on figures \ref{fig:scan_err_ampl1} and
\ref{fig:scan_err_ampl2}. Far from other pulses the reconstructed amplitudes are
within about \SI{\pm1}{\percent} around the true value. When pulses are
closer than \SI{\pm8}{\nano\second}, the errors in the reconstructed amplitudes
become more significant. The problems with the reconstructed amplitudes are severe
for pulses located close to pulses with significantly bigger amplitudes.

\section{Conclusions}
The proposed algorithm for deconvolution based reconstruction for the PADME
veto system is evaluated using artificially generated data resambling real
experimental condidions. The reconstructed times and amplitudes closly match their
original values for single pulses, but noticable deviation when
the pulses are heavily overlapped i.e. closer than \(\approx
\SI{20}{\nano\second}\).  Two pulses become indestinguishable if their
separation is below \SI{8}{\nano\second}.

The \SI{\pm8}{\nano\second} (\SI{\pm20}{samples})
veto window  applied to each reconstructed particle directly increase the
inefficiency of the experiment. Although the deocnvolution based reconstruction
using the estimated impulse response function outperforms the currently used
topology based method, means to reduce the veto window are desirable and will
be subject to future investigations.

The algorithms for spike detection and baseline drift correction are important
preprocessing steps for the impulse response function estimation, but are,
however, independent of the proposed deconvolution algorithm.  The suggested
algorithms are computationally cheap, which makes them suitable for online
data processing and for analysing large volums of data.

\section*{Acknowledgements}
This work is supported by MUCCA, CHIST-ERA-19-XAI-009 / BG-NSF KP-06-D002/4
from 15.12.2020 and funded by
The Bulgarian national program ``Young scientists and postdocs'' RMS 577/17.08.2018 and by
Sofia University, Bulgaria.

The PADME collaboration is acknowledged for the ability to test the developed
methods on PADME experimental data and for the provision of the physics case
which was a main inspiration for this development.


\begin{thebibliography}{9}

  \bibitem{ref:padme_ven} \href{http://dx.doi.org/10.1051/epjconf/20159601025}{Raggi M., Kozhuharov V., Valente P. The PADME experiment at LNF, {\it EPJ Web of Conferences} (2015) {\bf 96}.}
  \bibitem{ref:hindawi_ven} \href{https://doi.org/10.1155/2014/959802}{Raggi M, Kozhuharov V, Proposal to Search for a Dark Photon in Positron on Target Collisions at DA$\Phi$NE Linac", {\it Advances in High Energy Physics}, vol. {\bf 2014}, Article ID 959802, 14 pages, 2014.}
  \bibitem{ref:veto_prototype} \href{https://doi.org/10.1109/TNS.2018.2822724}{F. Ferrarotto et al., Performance of the Prototype of the Charged-Particle Veto System of the PADME Experiment, IEEE Transactions on Nuclear Science, vol. 65, no. 8, pp. 2029-2035, Aug. 2018.}
  \bibitem{ref:sipm_controller}
    \href{http://dx.doi.org/10.1088/1742-6596/898/3/032024}{S. Bertelli, et
    al., Design and performance of the front-end electronics of the charged
    particle detectors of PADME experiment, 2024 JINST 19 C01051}
  \bibitem{ref:emanuele_daq} \href{http://dx.doi.org/10.1088/1742-6596/898/3/032024}{E Leonardi et al.  2017 {\it J. Phys.}: Conf. Ser. {\bf 898} 032024.}
  \bibitem{ref:v_yordanov} \href{http://dx.doi.org/10.1016/j.nima.2015.07.040}{V. Jordanov, Unfolding-synthesis technique for digital pulse processing. Part 1: Unfolding, NIM-A, Volume 805, 2016, Pages 63-71 }
  \bibitem{ref:sac} \href{https://doi.org/10.1016/j.nima.2018.12.035} {A.  Frankenthal, et al., Characterization and performance of PADME’s Cherenkov-based small-angle calorimeter, NIM-A, Volume 919, 2019, Pages 89-97, }
  \bibitem{ref:ecal} \href{https://doi.org/10.1088/1748-0221/15/10/T10003}{P. Albicocco et al, The International School for Advanced Studies (SISSA), find out more Characterisation and performance of the PADME electromagnetic calorimeter, 2020 JINST 15 T10003 }
  \bibitem{ref:drs4_calibration} \href{https://doi.org/10.1088/1742-6596/2255/1/012008}{Georgi Georgiev for the PADME collaboration 2022 J. Phys.: Conf. Ser. 2255 012008, doi: 10.1088/1742-6596/2255/1/012008}
  \bibitem{ref:sparce_deconvolution} {Selesnick I., Sparce deconvolution (an MM algorithm), Polytechnic Institute of New York University, (2014)}

\end{thebibliography}
\end{document}